\documentclass[11pt]{revtex4}
\usepackage{bm}  % bold math
\usepackage{graphicx}
\usepackage{amssymb}
\usepackage{color}
\usepackage{amscd}
\usepackage{epsfig}
\newcommand{\vmean}{\overline{v}}

\newcommand{\la}{\langle}
\newcommand{\ra}{\rangle}

\newcommand{\dirl}{{i}}   % parallel index
\newcommand{\dirn}{{\ell}}         % layer (transverse) index
\newcommand{\dirm}{{j}}   % parallel nbr
\newcommand{\dirp}{{m}}            % transverse nbr index

\begin{document}

\title{Depinning and plasticity of driven disordered lattices}
\author{M. Cristina Marchetti}
\affiliation{Physics Department, Syracuse University, Syracuse, NY
13244, USA }
\date{\today}
\begin{abstract}
\noindent Invited lecture presented at the XIX Sitges Conference
on \emph{Jamming, Yielding, and Irreversible Deformations in
Condensed Matter}, Sitges, Barcelona, Spain, June 14-18, 2004.
\end{abstract}
\maketitle

\section{Introduction}
\label{Sec: Introduction}

Nonequilibrium transitions from stuck to flowing phases underlie
the physics of a wide range of physical phenomena. In a first
class of systems the onset of a stuck or frozen state occurs as a
result of intrinsic dynamical constraints, due to interactions or
crowding, and is usually referred to as \emph{jamming}
\cite{Liu1998}. Familiar examples are supercooled liquids that
become a glasses upon lowering the temperature,  colloidal
suspensions that undergo a glass transition due to crowding upon
increasing the density or the pressure, foams and granular
materials that jam under shear, arrays of dislocations in solids
that jam under an applied load. In a second class of systems the
transition to a stuck state is due to external constraints, such
as the coupling to quenched disorder (pinning centers from
material defects in vortex lattices, optical traps in colloids,
etc.), and is denoted as \emph{pinning} \cite{Fisher98}. Both
classes of systems can be driven in and out of glassy states by
tuning not only temperature, density or disorder strength, but
also an applied external force. The external drive may be a shear
stress in conventional glasses or simply a uniform applied force
in systems with extrinsic quenched disorder, where even a uniform
translation of the system relative to the fixed impurities
represents a nontrivial perturbation. Vortex lattices in
superconductors \cite{Blatter1994} and charge density waves (CDWs)
in metals \cite{Gruner1988} can be driven in and out of stuck
glassy states by a uniform external current or electric field,
respectively. As recognized recently in the context of jamming,
the external drive plays a role not unlike that of temperature in
driving the system to explore metastable configurations and should
be included as an axis in a complete phase diagram.

In this lectures I will focus on zero-temperature depinning
transitions of interacting condensed matter systems that
spontaneously order in periodic structures and are driven over
quenched disorder. The prototype examples are vortex lattices in
type-II superconductors \cite{Giamarchi2002} and charge density
waves in anisotropic metals \cite{DSF85}. Other examples include
Wigner crystals of two dimensional electrons in a magnetic field
moving under an applied voltage \cite{Andrei1988}, lattices of
magnetic bubbles  moving under an applied magnetic field gradient
\cite{Seshadri1992}, and many others. In general these systems
form a lattice inside a solid matrix, provided by the
superconducting or conducting material and are subject to pinning
by random impurities. The statics of such disordered lattices have
been studied extensively \cite{Giamarchi2002}. One crucial feature
that distinguishes the problem from that of disordered interfaces
is that the pinning force experienced by the periodic structure is
itself periodic, although with random amplitude and phase
\cite{GL1994&1995}. As a result, although disorder always destroys
true long-range translational order and yields glassy phases with
many metastable states and diverging energy barriers between these
states,  the precise nature of the glassy state depends crucially
on disorder strength. At weak disorder the system, although
glassy, retains topological order (the resulting phase has been
named Bragg glass in the context of vortex lattices)
\cite{GL1994&1995}. Topological defects proliferate only above
some characteristic disorder strength, where a topologically
disordered glass is formed.

The driven dynamics of disordered periodic structures have been
studied extensively by modeling the system as an overdamped
elastic medium that can be deformed by disorder, but is not
allowed to tear, that is by neglecting the possible formation of
topological defects due to the competition of elasticity, disorder
and drive. This model, first studied in the context of charge
density waves, exhibits a nonequilibrium phase transition from a
pinned  to a \emph{unique} sliding state at a critical value $F_T$
of the driving force. This nonequilibrium transition displays
universal critical behavior as in equilibrium {\it continuous}
transitions, with the medium's mean velocity $v$ acting as an
order parameter \cite{DSF85,NF92,Fisher98}. While the overdamped
elastic medium model may seem adequate to describe the dynamics of
driven Bragg glasses, many experiments and simulations of driven
systems have shown clearly that topological defects proliferate in
the driven medium even for moderate disorder strengths
\cite{Bhattacharya93,Nori96,Tonomura99,Troyanovski99,Maeda02}. The
dynamics near depinning  becomes spatially and temporally
inhomogeneous, with coexisting moving and pinned degrees of
freedom. This regime has been referred to as plastic flow and may
be associated with memory effects and even hysteresis in the
macroscopic response.

The goal of the present lectures is to describe coarse-grained
models of driven extended systems  that can lead to
history-dependent dynamics. Such models can be grouped in two
classes. In the first class the displacement of the driven medium
from some undeformed reference configuration remains
single-valued, as appropriate for systems without topological
defects, but the interactions are modified to incorporate
non-elastic restoring forces
\cite{littlewood88,levy92,levy94,MMP2000,SF01,MMSS2003}. In the
second class of models topological defects are explicitly allowed
by removing the constraint of single-valued displacements
\cite{strogatz88&89,NV97,SSMM04}. Here we will focus on the first
class and specifically consider driven periodic media with a
linear stress-strain relation, where the stress transfer between
displacements of different parts of the medium is nonmonotonic in
time and describes viscous-type slip of neighboring degrees of
freedom. A general model of this type  that encompasses many of
the models discussed in the literature was proposed recently by us
\cite{MMP2000,MCMKD02,MMSS2003}. Here slips between neighboring
degrees of freedom are described as viscous force, that allows a
moving portion of the medium to overshoot a static configuration
before relaxing back to it. It is shown below that such viscous
coupling can be considered an effective way of incorporating the
presence of topological defects in the driven medium. Related
models have also been used to incorporate the effect of inertia or
elastic stress overshoot in crack propagation in solids
\cite{SF01,SF02}. The precise connection between the two classes
of models has been discussed in Ref.~\cite{MCMPramana}.

In Section \ref{Sec: SingleParticle} we review the simplest
example of depinning transition, obtained when non-interacting
particles are driven through a periodic pinning potential. By
contrasting the case of periodic and non-periodic pinning, we
stress  that care must be used in the definition of the mean
velocity of the system. In Section \ref{Sec: ExtendedMedium}, we
first describe the generic coarse-grained model of a driven
elastic medium that exhibits a \emph{continuous} depinning
transition as a function of the driving force from a static to a
\emph{unique} sliding state.  Next we introduce an anisotropic
visco-elastic model as a generic model of a periodic system driven
through strong disorder.  The model considers coarse-grained
degrees of freedom that can slip relative to each other in the
directions transverse to the mean motion, due to the presence of
small scale defects (phase slips, dislocations, grain boundaries)
at their boundaries, but remain elastically coupled in the
longitudinal directions. The slip interactions are modeled as
viscous couplings and a detailed physical motivation for this
choice is given in section \ref{SUBSEC:viscous_couplings}. Most of
our current results for these type of models are for the
mean-field limit and are presented in Section
\ref{SEC:mean_field}. The studies carried out so far for
finite-range interactions suggest that the mean-field theory
described here may give the correct topology for the phase
diagram, although there will of course be corrections to the
critical behavior in finite dimensions \cite{Bety_unp}. Finally,
we conclude in Section \ref{SEC:other_models} by discussing the
relation to other models described in the literature and the
connection to experiments.

\section{Depinning of Noninteracting Particles}
\label{Sec: SingleParticle} It is instructive to begin with the
problem of a single particle driven through a \emph{periodic}
pinning potential as the simplest illustration of driven
depinning. Assuming overdamped dynamics, the equation of motion
for the position $x$ of the particle is
\begin{equation}\label{single_per}
\zeta\frac{dx}{dt}=F+hY(x)\;,
\end{equation}
where $\zeta$ is a friction coefficient (in the following we
choose our units of time so that $\zeta=1$), $F$ is the external
drive and $Y(x)=Y(x+n)$, with $n$ an integer, is a periodic
function of period $1$. For simplicity we choose a piecewise
linear pinning force, corresponding to $Y(x)=(1/2-x)$, for $0\leq
x\leq 1$. In this case a periodic solution of
Eq.~(\ref{single_per}) is obtained immediately in terms of the
time $T$ needed to traverse a potential well, or period.
Introducing an arbitrary time $t_J$ such that if $x(t_J)=n$, then
$x(t_J+T)=n+1$, the particle position for $t_J+nT\leq t\leq
t_J+(n+1)T$ is
\begin{equation}
\label{single_soltn}
x(t)=n+\frac{1-e^{-h(t-t_J-nT)}}{1-e^{-hT}}\;,
\end{equation}
where $T$ is given by
\begin{equation}
\label{period} T(h)=\frac{1}{h}\ln\bigg(\frac{2F+h}{2F-h}\bigg)\;,
\end{equation}
for $F>h/2$ and diverges for $F<h/2$. In other words if $F<h/2$
the particle never leaves the initial well, i.e., it is pinned.
The threshold force for depinning is then $F_c=h/2$. In the
sliding state the mean velocity is defined as the average of the
instantaneous velocity $v(t)=\frac{dx}{dt}$ over the arbitrary
initial time $t_J$. This gives
\begin{equation}
\label{vmean_single} \vmean\equiv\langle
v\rangle_{t_J}=\int_{t-(n+1)T}^{t-nT}\frac{dt_J}{T}~
v(t)=\frac{1}{T}\;.
\end{equation}
This definition naturally identifies the mean velocity of the
particle with the inverse of the period. The logarithmic behavior
of  $\vmean$ near threshold, $\vmean\sim -1/ln(F_c-F)$, is
peculiar to a discontinuous pinning force. For an arbitrary
pinning force $Y(X)$ the period $T$ is
\begin{equation}
\label{period_int} T=\int_0^1 dx \frac{1}{F+hY(x)}\;,
\end{equation}
and can be evaluated analytically for various forces. For
instance, for a sinusoidal pinning force, $Y(x)=\sin(2\pi x)$, one
finds $T=(F^2-h^2)^{-1/2}$, which gives $\vmean\sim(F-F_c)^{1/2}$
near threshold, a generic behavior for continuous pinning forces.

The main focus of the remainder of this paper will be on the
modeling of extended driven systems as collections of interacting
degrees of freedom. It will then be important to distinguish two
cases. For extended systems that are periodic, such as charge
density waves and vortex lattices, the pinning potential is itself
periodic as each degrees of freedom sees the same disorder after
advancing one lattice constant. For non-periodic systems, such as
interfaces, each degree of freedom moves through a random array of
defects. When interactions are neglected, an extended
\emph{periodic} system moving through a periodic random pinning
potential can be modeled as a collection of $N$ non-interacting
particles, where each particle sees its own periodic pinning
potential. The pinning potentials seen by different particles may
differ in height and be randomly shifted relative to each other,
as sketched in Fig. 1. The equation of motion for the $i$-th
particle at position $x_i$ is then
\begin{equation}
\label{many_per} \frac{dx_i}{dt}=F+h_iY(x_i+\gamma_i)\;,
\end{equation}
where $\gamma_i$ are random phases uniformly distributed in
$[0,1)$ and the pinning strengths $h_i$ are drawn independently
from a distribution $\rho(h_i)$. Since the displacements $x_i$ are
decoupled, they can be indexed by their disorder parameters
$\gamma$ and $h$ instead of their spatial label $i$, i.e.,
$x_i(t;\gamma_i,h_i)\rightarrow x(t;\gamma,h)$. The mean velocity
of the many-particle system can then be written as an average over
the random phases and pinning strengths,
\begin{eqnarray}
\vmean=\frac{1}{N}\sum_i v_i
&=&\la v(t;\gamma,h)\ra_{\gamma, h}\nonumber\\
&=& \int dh ~\rho(h)\int_0^1d\gamma ~v(t;\gamma,h)\;,
\end{eqnarray}
where $v(t;\gamma,h)=\frac{dx(t;\gamma,h)}{dt}$.  The average over
the random phase of each degree of freedom is equivalent to the
average over the random time shift $t_J$ described for the
single-particle case and yields $\int_0^1d\gamma~
v(t;\gamma,h)=1/T(h)$, with $T(h)$ the period of each particle
given in Eq.~(\ref{period}). The mean velocity is then
\begin{equation}
\label{vmean_per}
\vmean=\bigg\langle\frac{1}{T(h)}\bigg\rangle_h\;,
\end{equation}
where $\langle ...\rangle_h=\int dh ...\rho(h)$ denotes the
average over the barrier height distribution. For distributions
$\rho(h)$ that have support at $h=0$, a system of
\emph{noninteracting} particles with periodic pinning depins at
$F=0$, as there are always some particles experiencing zero
pinning force.

\begin{figure}
\centering
\includegraphics[width=6.cm]{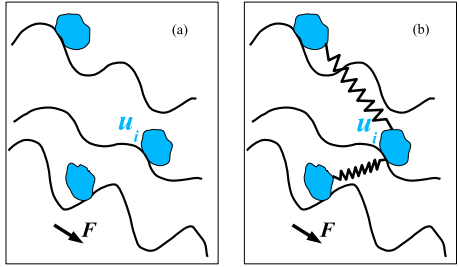}
\caption{\label{fig:1Dcartoon} (a) Sketch of noninteracting
degrees of freedom  driven over a random periodic pinning
potential in one dimension.  Spatial coordinates have been
discretized so that degrees of freedom are labelled by an index
$i$. In (b) the case where each degree of freedom interacts
elastically with its neighbors is shown. This is a discretized
one-dimensional realization of the elastic medium model described
by Eq.~(\ref{elastic_model}) below.}
\end{figure}

A different single-particle problem that has been discussed in the
literature is that of a particle moving through a \emph{random
(non-periodic)} array of defects \cite{Zapperi2000}. The defects
can be described as pinning potential wells centered at random
positions and/or with random well heights. To make contact with
the periodic case we consider a particle moving through a
succession of evenly spaced pinning potential wells of random
heights. The equation of motion is
\begin{equation}
\label{single_nonper}
\partial_t x=F+\sum_{p=0}^{N_p-1}h_pY(x-p)\;,
\end{equation}
where $N_p$ is the total number of pinning centers and the pinning
strengths $h_p$ are drawn independently from a distribution
$\rho(h_p)$. Choosing again the piecewise-linear pinning force,
the time to traverse the $p$-th well is simply $T(h_p)$, with $T$
given by Eq.~(\ref{period}). The mean velocity of the particle is
defined as the total distance travelled divided by the total time
and is given by
\begin{equation}
\label{vmean_nonper} \vmean=\frac{N_p}{\sum_p T(h_p)}\equiv
\frac{1}{\langle T(h)\rangle_h}\;.
\end{equation}
In this case, unless the distribution $\rho(h)$ is bounded from
above, there is always a finite probability that the particle will
encounter a sufficiently deep potential well to get pinned.
Therefore for unbounded $\rho(h)$ the particle is always pinned in
the thermodynamic limit. If $\rho(h)$ is bounded from above by a
maximum pinning strength $h_{\rm max}$, this value also represents
the depinning threshold. Finally, the case of many noninteracting
particles driven through a random array of defects is equivalent
to that of a single particle, as the mean velocity of each
particle can be calculated independently. The mean velocity of the
system is then again given by Eq.~(\ref{vmean_nonper}).

\section{Depinning of an Extended Medium}
\label{Sec: ExtendedMedium} We consider a $d$-dimensional periodic
structure driven along one of its symmetry directions, chosen as
the $x$ direction. The continuum equations for such a driven
lattice within the elastic approximation were derived by various
authors by a rigorous coarse-graining procedure of the microscopic
dynamics \cite{BMR1998,SV1998,GL1998}. Assuming overdamped
microscopic dynamics, the equation for the local deformation ${\bf
u}({\bf r},t)$ of the medium (in the laboratory frame) from an
undeformed reference state is written by balancing all the forces
acting on each portion of the system as  \cite{convective_note}
\begin{equation}
\label{eqmotion}
\partial_tu_i=\partial_j\sigma_{ij}+F\delta_{ix}+F_{pi}({\bf
r},{\bf u})\;,
\end{equation}
where $\sigma_{ij}$ is the stress tensor due to interactions among
neighboring degrees of freedom, $F$ is the driving force and ${\bf
F}_p$ is the periodic pinning force.  The periodicity of the
pinning force, which contains Fourier components at all the
reciprocal lattice vectors of the lattice, arises from the
coupling to the density of the driven lattice.

\subsection{Elastic model} \label{SUBSEC:elastic_model} For
conventional short-ranged elasticity the stress tensor is
\begin{equation}
\label{stress} \sigma_{ij}^{\rm
el}=2c_{66}u_{ij}+\delta_{ij}(c_{11}-c_{66})u_{kk}\;,
\end{equation}
where $c_{11}$ and $c_{66}$ are the compressional and shear moduli
of the driven lattice, respectively, and
$u_{ij}=\frac{1}{2}(\partial_iu_j+\partial_ju_i)$ is the strain
tensor. It was shown in Ref.~\cite{BMR1998} that deformations of
the driven lattice along the direction of the driving force grow
without bound due to large transverse shear stresses that generate
unbounded strains responsible for dislocation unbinding. For this
reason, we focus here on the dynamics of a scalar field
$u_x(x,{\bf y},t)\equiv u({\bf r},t)$, with ${\bf r}=(x,{\bf y})$,
describing deformations of the driven lattice along the direction
of mean motion. The $d-1$-dimensional vector ${\bf y}$ denotes the
coordinates transverse to the direction of motion. Assuming
$c_{11}>>c_{66}$, we obtain a scalar model for the driven elastic
medium, given by
\begin{equation}
\label{elastic_model}
\partial_tu=c_{11}\partial_x^2u+c_{66}\nabla_{\bf
y}^2u+F+F_p({\bf r},u)\;,
\end{equation}
where $F_p$ denotes the $x$ component of the pinning force. For
simplicity we also consider a model that only retains the
component of the pinning force at the smallest reciprocal lattice
vector and choose our units of lengths so that the corresponding
period is 1. The pinning force is then taken of the form
\begin{equation}
\label{Fp} F_p({\bf r},u)=h({\bf r})Y\big(u({\bf r},t)-\gamma({\bf
r})\big)\;,
\end{equation}
where $Y(u)=Y(u+n)$ is a periodic function. The random pinning
strengths $h$ are drawn independently at every spatial point from
a distribution with zero mean and short-ranged correlations to be
prescribed below. The random phases $\gamma$ are spatially
uncorrelated and distributed uniformly in $[0,1)$.

The model of a driven overdamped elastic medium embodied by
Eq.~(\ref{elastic_model}) has been studied extensively both
analytically and numerically
\cite{DSF85,NF92,Fisher98,aamSIM,myersSIM}. It exhibits a
depinning transition at a critical value $F_T$ of the applied
force from a static to a \emph{unique} sliding state
\cite{AAMunique}. The depinning can be described as a continuous
equilibrium transition, with the mean velocity
$\overline{v}=\la\partial_t u\ra$ playing the role of the order
parameter, and universal critical behavior. The velocity vanishes
as $F_T$ is approached from above as
$\overline{v}\sim(F-F_T)^\beta$. The critical exponent $\beta$
depends only on the system dimensionality and was found to be
$\beta=1-\epsilon/6+{\cal O}(\epsilon^2)$ using a functional RG
expansion in $\epsilon=4-d$ \cite{NF92,DWC2004}.

\subsection{Viscoelastic model}
\label{SUBSEC:VE_model} Strong disorder can yield topological
defects in the driven lattice, making the elastic model
inapplicable \cite{coppersmith90&91,sncajm91}. In this case the
dynamics becomes inhomogeneous, with coexisting pinned and moving
regions \cite{ShiBerlinsky1991, Faleski1996}. The depinning
transition may be discontinuous (first order), possibly with
macroscopic hysteresis. Several mean-field models of driven
extended systems have been proposed
\cite{strogatz88&89,levy92,levy94,NV97,Fisher98,MMP2000,SF01} to
describe this inhomogeneous dynamics. Here we focus on a class of
models that retains a single-valued displacement field and a
linear stress-strain relation, but assumes that the presence of
topological defects can be effectively incorporated at large
scales by a non-instantaneous stress transfer that couples to
 gradients of the local velocity (rather than
displacement). More precisely, we consider an anisotropic model of
coarse-grained degrees of freedom that can slip relative to each
other in at least one of the directions transverse to the mean
motion, due to the presence of small scale defects (phase slips,
dislocations, grain boundaries) at their boundaries, but remain
elastically coupled in the longitudinal directions
\cite{MMSS2003}.  This model incorporates the anisotropy of the
sliding state in the plastic flow region that results either from
flow along coupled channels oriented in the direction of the drive
(e.g., as in the moving smectic phase \cite{BMCMR97}) or in
layered materials such as the high-$T_c$ cuprate superconductors.
It also encompasses several of the models discussed in the
literature.

\begin{figure}
\centering
\includegraphics[width=8.cm]{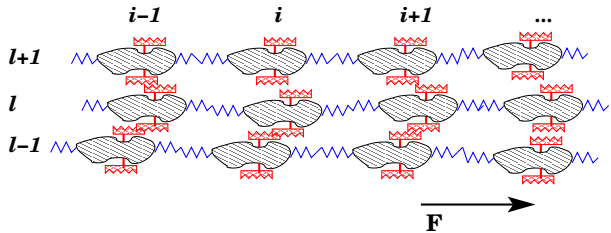}
\caption{\label{fig:cartoon} A two-dimensional realization of the
anisotropic driven medium described in the text. Spatial
coordinates have been discretized in the figure so that degrees of
freedom are labelled by indices $\dirn$ and $\dirl$, respectively
transverse and longitudinal to the direction of the driving force,
$F$. Each degree of freedom interacts with its neighbors via
elastic couplings in the longitudinal direction and via viscous or
similar slip couplings in the transverse direction.}
\end{figure}

For generality, consider a $d=d_\parallel+d_\perp$-dimensional
medium composed of degrees of freedom that are coupled elastically
in $d_\parallel$ direction and can slip relative to each other in
the remaining $d_\perp$ directions. The axis $x$ along which the
driving force is applied is along one of the $d_\parallel$
directions. The equation of motion for the displacement $u({\bf
r}_\parallel,{\bf r}_\perp,t)$ is given by
\begin{equation}
\label{VE_model}
\partial_tu=K\nabla_\parallel^2u+\eta\nabla_\perp^2v+F+F_p({\bf
r},u);,
\end{equation}
with $v=\partial_tu$ the local velocity. This model will be
referred to as the visco-elastic (VE) model as it incorporates
elastic couplings of strength $K$ in $d_\parallel$ directions and
viscous couplings of strength controlled by a shear viscosity
$\eta$ in the remaining $d_\perp$ directions. A two-dimensional
cartoon of this anisotropic model is shown in
Fig.~\ref{fig:cartoon}.

For $\eta=0$ (or $d_\perp=0$) the VE model reduces to the elastic
model (but with isotropic elasticity) of
Eq.~(\ref{elastic_model}).  Conversely, for $K=0$ (or
$d_\parallel=0$) Eq.~(\ref{VE_model}) reduces to the purely
viscous model studied earlier by us \cite{MMP2000,MCMKD02}. For
any distribution of pinning strengths with support at $h=0$, the
purely viscous model has zero threshold for depinning, but it does
exhibit a critical point separating regions of unique and
multivalued solutions for the mean velocity. In the VE model
($\eta\not=0$ and $K\not=0$) even when fluid-like shear takes
place, particle conservation gives a sharp depinning transition in
flow  along the channels. Furthermore, as shown below, the model
has a sharp mean-field tricritical point separating a region of
parameters where depinning is continuous, in the universality
class of elastic depinning, from one where depinning become
discontinuous and hysteretic.

 It is important to stress that the VE model still assumes
\emph{overdamped microscopic dynamics}. Velocity or viscous
couplings  can appear generically in the large-scale equations of
motion upon coarse-graining the microscopic dynamics of a
dissipative medium. In fact, next we show that viscous couplings
indeed represent an effective way of incorporating the local
dissipation due to the presence of topological defects.

\subsection{Viscoelastic coupling as an effective description of
topological defects} \label{SUBSEC:viscous_couplings}The goal of
this section is to provide some justification to the  anisotropic
VE model as an effective description of topological defects in a
driven lattice. To this purpose we consider a two dimensional
medium and take advantage of the continuum equations developed
many years ago by Zippelius et al. \cite{ZHN1980} to describe the
time-dependent properties of two-dimensional solids near melting.
These authors combined the equations of free dislocation motion
with solid hydrodynamics to construct a semimicroscopic dynamical
model of a solid with free dislocations. They further showed that
the dynamics of such of a "heavily dislocated solid" (an elastic
medium with an equilibrium concentration of free dislocations) is
identical to that of the hexatic phase obtained when a
two-dimensional solid melts via the unbinding of dislocations
\cite{Nelson_Domb1983}.  More recently we \cite{MCMKS02}
reconsidered the dynamical equations for the "heavily dislocated
solid" of Ref. \cite{ZHN1980} and showed that they can be recast
in the form of the phenomenological equations of a viscoelastic
fluid (with hexatic order) introduced many years ago by Maxwell
\cite{boonyip80}. In the presence of free dislocations the local
stresses in the medium have contributions from both elastic
stresses and defect motion. The latter couple again to the the
local strains which control the defect dynamics. By eliminating
the defect degrees of freedom, one obtains a linear, although
nonlocal, relation between strain and stress, given by
\cite{nonlocality_note}
\begin{eqnarray}
\label{sigmaVE} \sigma^{\rm VE}_{ij}({\bf r},t)=&&\delta_{ij}~
c_L~u_{kk}({\bf r},t)+\delta_{ij}(c_{11}-c_L)\int_{-\infty}^tdt'
e^{-(t-t')/\tau_b}~v_{kk}({\bf r},t')\nonumber\\
&&+2c_{66}\int_{-\infty}^tdt'e^{-(t-t')/\tau_s}\big[v_{ij}({\bf
r},t')-\frac{1}{2}\delta_{ij}v_{kk}({\bf r},t')\big]\;,
\end{eqnarray}
where $v_{ij}=\frac{1}{2}(\partial_iv_j+\partial_jv_i$ and the
velocity ${\bf v}$ is defined here in terms of the momentum
density ${\bf g}$ as ${\bf v}={\bf g}/\rho_0$, with $\rho_0$ the
equilibrium mass density of the medium. Also in
Eq.~(\ref{sigmaVE}) $c_L$ is the compressional modulus of the
liquid and $\tau_b\approx(c_{11}\mu_d^cn_fa_0^2)^{-1}$ and
$\tau_s\approx(c_{66}\mu_d^gn_fa_0^2)^{-1}$ are the compressional
and shear relaxation times, with $\mu_d^{g,c}$ the dislocation
glide and climb mobility, respectively. Of course in the presence
of dislocations the displacement $u$ is no longer single-valued
(although the strain $u_{ij}$ remains single-valued and
continuous) and $\partial_tu\not=v$ due to both the motion of
vacancy/interstitial defects and of dislocations. The
phenomenological Maxwell model of viscoelasticity is obtained by
assuming that $\partial_tu=v$ even in the presence of
dislocations. Then for $t<<\tau_s,\tau_b$ the viscoelastic stress
$\sigma^{\rm VE}({\bf r},t)$  reduces to the familiar elastic
stress tensor given in Eq.~(\ref{stress}),
\begin{equation}
\sigma^{\rm VE}({\bf r},t<<\tau_s,\tau_b)\approx \sigma^{\rm
el}_{ij}\;. \end{equation}
Conversely for $t>>\tau_s,\tau_b$ one obtains
\begin{equation}
\label{sigma_fluid} \sigma^{\rm VE}_{ij}({\bf
r},t>>\tau_s,\tau_b)\approx
\delta_{ij}~c_Lu_{kk}+\delta_{ij}(\eta_b+\eta)v_{kk}+2\eta
v_{ij}\;,
\end{equation}
which describes stresses in a viscous fluid of shear viscosity
$\eta=c_{66}\tau_s$ and bulk viscosity
$\eta_b=(c_{11}-c_{L})\tau_b$. The first term on the right hand
side of Eq.~(\ref{sigma_fluid}) is the pressure and incorporates
the fact that even a liquid has a nonzero long-wavelength
compressional elasticity, which is associated with density
conservation. As we will see below this terms plays a crucial role
in controlling the physics of depinning of a viscoelastic medium.
The Maxwell viscoelastic fluid has solid-like shear rigidity at
high frequency, but flows like a fluid at low frequency. Since the
relaxation times $\tau_s$ and $\tau_b$ are inversely proportional
to the density $n_f$ of free dislocations, the Maxwell model
behaves as a continuum elastic medium on all time scales when
$n_f\rightarrow 0$ and as a viscous fluid when $n_fa_0^2\sim 1$.

Dislocation climb is much slower than dislocation glide
($\mu_d^c<<\mu_d^g$), resulting in $\tau_b>>\tau_s$. We therefore
assume that the response to compressional deformations is
instantaneous on all time scales, but retain a viscoelastic
response to shear deformations. Letting $\tau_b\rightarrow\infty$,
we find
\begin{eqnarray}
\sigma_{ij}^{\rm VE}({\bf r},t)\approx \delta_{ij}~c_{11}u_{kk}(t)
+2c_{66}\int_{-\infty}^tdt'e^{-(t-t')/\tau_s}\big[v_{ij}(t')-\frac{\delta_{ij}}{2}v_{kk}(t')\big]\;.
\end{eqnarray}

We now turn to the case of interest here, where topological
defects are generated in a an extended medium driven through
quenched disorder. In this case the medium has no low frequency
shear modulus, but particle conservation still requires long
wavelength elastic restoring forces to \emph{compressional}
deformations. On the other hand, the  number of topological
defects is not fixed as dislocations are continuously generated
and annihilated by the interplay of elasticity, disorder and drive
\cite{ShiBerlinsky1991,KoshelevVinokur1994,Faleski1996}.
Furthermore, unbound dislocations can be pinned by disorder and do
not equilibrate with the lattice. In the plastic region near
depinning the dynamics remains very inhomogeneous and fluid-like
and the pinning of dislocations by quenched disorder is not
sufficient to restore the long wavelength shear-stiffness of the
medium. For this reason we propose to describe the effect of
topological defects near depinning by replacing elastic
\emph{shear} stresses by viscoelastic ones, while retaining
elastic \emph{compressional} forces. Of course the resulting model
that assumes a fixed density of dislocations becomes inapplicable
at large driving forces where dislocations heal as the lattice
reorders. For the case of interest here of a scalar model
describing only deformations along the direction of motion, the
viscoelastic model of a driven disordered medium is
\begin{equation}
\label{full_VE}
\partial_tu=c_{11}\partial_x^2u+c_{66}\int_{-\infty}^tdt'
e^{-(t-t')/\tau_s}\partial_y^2v(t')+F+h({\bf r})Y(u-\gamma({\bf
r}))\;,
\end{equation}
with $v=\partial_tu$. This model naturally incorporates the
anisotropy and channel-like structure of the driven medium, where
\emph{shear} deformations due to gradients in the displacement in
the directions transverse to the mean motion ($\partial_y
u\not=0$) are most effective at generating the large stresses
responsible for the unbinding of topological defects. It is
instructive to note that due to the exponential form of stress
relaxation the integro-differential equation (\ref{full_VE}) is
equivalent to a second order differential equation for the
displacement,
\begin{equation}
\label{utwodots} \tau_s\partial_t^2u+\gamma_{\rm
eff}\partial_tu=c_{11}\partial_x^2u+\eta\partial_y^2v+F+h({\bf
r})Y(u-\gamma({\bf r}))\;,
\end{equation}
with $\gamma_{\rm eff}$ an effective friction \cite{MMP2000}. In
other words the effect of a finite density of dislocations in the
driven lattice yields  "inertial effects" on a scale controlled by
the time $\tau_s\sim1/n_f$. The purely viscous model obtained from
Eq.~(\ref{utwodots}) with $c_{11}=0$ was analyzed in detail in
Ref. \cite{MCMKD02} where it was shown that if $\tau_s$ and
$\eta=c_{66}\tau_s$ are tuned independently, then $\tau_s$ is a
strongly irrelevant parameter in the RG sense. This allows us to
consider a simplified form of the equation for the driven medium
obtained  from  Eq.~(\ref{utwodots}) with $\tau_s=0$, but
$\eta=c_{66}\tau_s$ finite, leading to the general anisotropic
viscoelastic model introduced in Eq.~(\ref{VE_model}).

\section{Mean-field solution}
\label{SEC:mean_field} The mean-field approximation for the VE
model is obtained in the limit of infinite-range elastic and
viscous interactions. To set up the mean field theory, it is
convenient to discretize space in both the transverse and
longitudinal directions, using integer vectors $\dirl$ for the
$d_\parallel$-dimensional intra-layer index and $\dirn$ for the
$d_\perp$-dimensional layer index.  The local displacement along
the direction of motion is $u_\dirn^\dirl(t)$.  Its dynamics is
governed by the equation,
\begin{equation}
\label{VEmodel_discrete}
\partial_t u_\dirn^\dirl=\sum_{\la j\ra}K_{ij}(u_\dirn^\dirl-u_\dirn^\dirm)
   +\sum_{\la m\ra}
   \eta_{\ell m}[\dot{u}_\dirn^\dirl-\dot{u}_\dirm^\dirp]
    +F+h_\dirn^\dirl Y(u_\dirn^\dirl-\gamma_\dirn^\dirl)\;,
\end{equation}
where the dot denotes a time derivative  and $\la \dirm\ra$ ($\la
m\ra$) ranges over sites $\dirm$ ($m$) that are nearest neighbor
to $\dirl$ ($\ell$). The random pinning strengths $h_\dirn^\dirl$
are chosen independently with probability distribution
$\rho(h_\dirn^\dirl)$ and the $\gamma_\dirn^\dirl$ are distributed
uniformly and independently in $[0,1)$. For a system with
$N=N_\parallel\times N_\perp$ sites, \emph{one} mean field theory
is obtained by assuming that all sites are coupled with uniform
strength, both within a given channel and with other channels.
Each discrete displacement then couples to all others only through
the mean velocity, $\overline{v}=N^{-1}\sum\dot{u}_\dirn^\dirl$,
and the mean displacement,
$\overline{u}=N^{-1}\sum{u}_{\dirn}^\dirl$.  We assume that the
disorder is isotropic and the system is self averaging and look
for solutions moving with stationary velocity:
$\overline{u}=\overline{v}t$. Since all displacements $u$ are
coupled, they can now be indexed by their disorder parameters
$\gamma$ and $h$, rather than the spatial indices $\dirl$ and
$\dirn$. The mean-field dynamics is governed by the equation
\begin{equation}
\label{MFT_VE} (1+\eta)\dot{u}(\gamma,h)= K\big(\vmean t-u\big)
    +F
   +\eta\vmean+hY(u-\gamma).
\end{equation}
The cases $K = 0$ and $K\not=0$ need to be discussed separately.

\subsection{Mean-field theory for viscous model: $K=0$,
$\eta\not=0$} \label{SUBSEC:MFT_viscous} 
\begin{figure}
\centering
\includegraphics[width=12.cm]{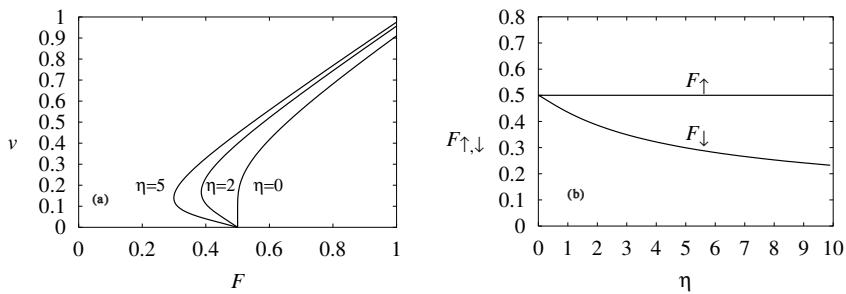}
\caption{\label{fig:viscous_fixed_h} (a) Velocity versus driving
force for the purely viscous model ($K=0$, $\eta\not=0$) with a
narrow distribution of pinning strength, $\rho(h)=\delta(h-1)$,
for $\eta=0,2,5$. There is a finite depinning threshold at
$F_T=1/2$. In $(b)$ the depinning and repinning forces
$F_\uparrow$ and $F_\downarrow$ are shown as  functions of
$\eta$.}
\end{figure}

\begin{figure}
\centering
\includegraphics[width=12.cm]{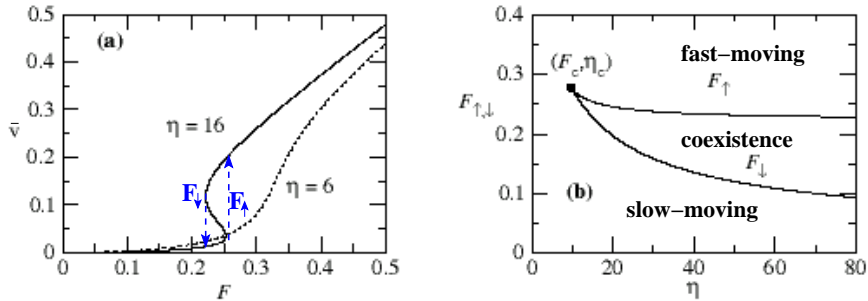}
\caption{\label{fig:viscous_broad_h} (a) Velocity versus driving
force for the purely viscous model ($K=0$, $\eta\not=0$) with a
broad distribution of pinning strength, $\rho(h)=e^{-h}$ for
$\eta=6,16$. In this case there are no stable static (pinned
states). The velocity is single valued for $\eta<\eta_c$ and
multi-valued for  $\eta>\eta_c$. In this case when $F$ is ramped
up from zero, the velocity jumps discontinuously at $F_\uparrow$
where the system goes from the "slow-moving" to the "fast-moving"
state. Here and below "coexistence" refers to multistability of
the solutions to the equations of motion. When $F$ is then ramped
down from within the fast-moving state the jump in $\vmean$ occurs
at the lower value $F_\downarrow$.  The forces $F_\downarrow$ and
$F_{\uparrow}$ become equal at the critical point, as shown in
frame (b).}
\end{figure}

When $K=0$, the mean
field equation becomes identical to that of a single particle
discussed in Section \ref{Sec: SingleParticle} driven by an
effective force $F+\eta\vmean$ (with friction $1+\eta$). In this
case different degrees of freedom  move at different velocities
even in the mean field limit. The mean field velocity is
determined by the self-consistency condition
$\vmean=\la\dot{u}\ra_{\gamma,h}$, where the average over the
random phases is equivalent to the average over the random times
shifts $t_J$ given in Eq.~(\ref{vmean_single}). For the case of a
piecewise linear pinning force using Eq.~(\ref{vmean_per})  we
find
\begin{equation}
\label{vmean_viscous}
\vmean=\frac{1}{1+\eta}\int
dh\rho(h)\frac{1}{T(h,F+\eta\vmean)}\;,
\end{equation}
with $T(h,F)$ given by Eq.~(\ref{period}). The mean velocity
obtained by self-consistent solution of Eq.~(\ref{vmean_viscous})
is shown in Figs.~\ref{fig:viscous_fixed_h} and
\ref{fig:viscous_broad_h} for two distributions of pinning
strengths.

For a narrow distribution, $\rho(h)=\delta(h-1)$, there is a
finite threshold $F_T=1/2$, independent of $\eta$. The velocity is
multivalued for any finite $\eta$. When the force is ramped up
adiabatically from the static state the system depins at
$F_\uparrow=F_T$. When the force is ramped down from the sliding
state, the system repins at the lower value $F_\downarrow(\eta)$.
The depinning and repinning forces are shown in
Fig.~\ref{fig:viscous_fixed_h}(b). The region where unique and
multivalued velocity solutions coexist  extend to $\eta=0$. For a
broad distribution with support at $h=0$, e.g., $\rho(h)=e^{-h}$,
the threshold for depinning is zero as some of the degrees of
freedom always experience zero pinning and start moving as soon as
a force is applied. There is a critical point at $(F_c,\eta_c)$.
For $\eta>\eta_c$ the analytical solution for $\vmean(F)$ is
multivalued, as shown in Fig.~\ref{fig:viscous_broad_h}. If the
force is ramped up adiabatically from zero at a fixed
$\eta>\eta_c$, the system depins discontinuously at
$F_\uparrow(\eta)$, while when the force is ramped down it repins
at the lower value $F_\downarrow(\eta)$, as shown in
Fig.~\ref{fig:viscous_broad_h}. The viscous model has also been
studied in finite dimensions by mapping it onto the nonequilibrium
random field Ising model (RFIM) \cite{MCMKD02}. In the mapping,
the local velocities correspond to spin degrees of freedom, the
driving force is the applied magnetic field and the mean velocity
maps onto the magnetization. The RFIM has a critical point
separating a region where the magnetization versus applied field
curve displays hysteresis with a discontinuous jump to a region
where there is no jump in the hysteresis curve
\cite{DahmenSethna1993,Silveira1999}. In the viscous model the
critical point separates a region where the velocity curve is
smooth and continuous from the region where the "depinning" (from
"slow-moving" to "fast-moving" states) is discontinuous and
hysteretic. The critical point is in the Ising universality class,
with an upper critical dimension $d_c=6$.

\subsection{Mean-field theory for VE model: $K\not=0$ and
$\eta\not=0$} \label{SUBSEC:MFT_viscous} When $K\ne 0$, all
degrees of freedom are coupled by a spring-like interaction (the
first term on the right hand side of Eq.~(\ref{MFT_VE})) to the
mean field $\overline{u}=\vmean t$ and cannot lag much behind each
other. This forces all the periods to be the same, independent of
$h$,  and yields a nonvanishing threshold for depinning. In this
case the mean field  velocity is determined by imposing $\la
u(t;\gamma,h)-\vmean t\ra_{\gamma,h}=0$.

It is useful to first review the case where $K\ne 0$ and $\eta=0$.
In this limit, Eq.~(\ref{MFT_VE}) reduces to the mean field theory
of a driven elastic medium worked out by Fisher and collaborators
\cite{NF92}. No moving solution exists above a finite threshold
force $F_T$. For the piecewise linear pinning force this is given
by
\begin{equation}
\label{FT} F_T=\la\frac{h^2}{2(K+h)}\ra_h\;. \end{equation}
For
$F>F_T$ there is a \emph{unique} moving solution that has  a
universal dependence on $F$ near $F_T$, where it vanishes as
$\vmean\sim(F-F_T)^\beta$. In mean-field the critical exponent
$\beta$ depends on the shape of the pinning force: $\beta=1$ for
the discontinuous piecewise linear force and $\beta=3/2$ for
generic smooth forces. Using a functional RG expansion in
$\epsilon=4-d$, Narayan and Fisher \cite{NF92} showed that the
discontinuous force captures a crucial intrinsic discontinuity of
the large scale, low-frequency dynamics, giving the general result
$\beta=1-\epsilon/6+{\cal O}(\epsilon^2)$, in reasonable agreement
with numerical simulations in two and three dimensions
\cite{myersSIM,aamSIM}. For simplicity and to reflect the
``jerkiness'' of the motion in finite-dimensional systems at low
velocities, we use piecewise linear pinning below.

When $\eta>0$ the nature of the depinning differs qualitatively
from the $\eta=0$ case, in that hysteresis in the dynamics can
take place. Again, no self-consistent moving solution exists for
$F<F_T$, with $F_T$ independent of $\eta$. Above threshold, both
unique and multi-valued moving solutions exist, depending on the
values of the parameters: $\eta$, $K$, and the shape of the
disorder distribution, $\rho(h)$. To obtain the mean field
solution in the sliding state, we examine the motion during one
period $T=1/\vmean$ during which the displacement advances by 1.
Eq.~(\ref{MFT_VE}) is is easily solved for $0\leq u\leq 1$ and
$\gamma=0$, with the result,
\begin{equation}
\label{MFsolution}
u(t;\gamma=0,h)=\frac{K\overline{v}t+F+\eta\overline{v}+h/2}{(1+\eta)\lambda}
  -\frac{K\overline{v}}{(1+\eta)\lambda^2}+Ae^{-\lambda t}\;,
\end{equation}
where $\lambda=(K+h)/(1+\eta)$. At long times, regardless of the
initial condition, $u(t)$ approaches a periodic function of period
$T=1/\vmean$ with jumps in its time derivative at times $t_J+nT$,
with $n$ an integer. The constant $A$ is determined by requiring
that if $u(t_J+nT)=n$, then $u(t_J+(n+1)T)=n+1$. Writing
$u(t;\gamma,h)=\vmean t+\tilde{u}$, it is easy to see that for an
arbitrary value of $\gamma$, the solution $\tilde{u}$ will have
the form $\tilde{u}=\tilde{u}(\overline{v}t-\gamma,h)$. The mean
velocity is then obtained from $\la
\tilde{u}(\overline{v}t-\gamma,h)\ra_{\gamma,h}=0$. Averaging
$\tilde{u}$ over $\gamma$ is equivalent to averaging $\tilde{u}$
for a fixed $\gamma$ over a time period, $T$, with the result,
\begin{eqnarray}
\langle\tilde{u}\rangle_\gamma & = & \int_{t_J+nT}^{t_J+(n+1)T}
\frac{dt}{T}~
  \tilde{u}(\vmean-\gamma t,h)\nonumber\\
&=& \frac{F+\eta\overline{v}}{K} -\frac{h^2}{2K(K+h)}
    -\frac{(K+2h)\overline{v}}{\lambda(K+h)}
    -\frac{h^2}{K(K+h)}\frac{1}{e^{\lambda/\overline{v}}-1}\;.
\end{eqnarray}
Finally, averaging over $h$ and using the consistency condition,
we obtain
\begin{eqnarray}
\label{MFvFApp} F-F_T=\overline{v}\big[1-M(\eta,K)\big]
  +\Big\langle\frac{h^2}{K(K+h)}\frac{1}{e^{(K+h)/(1+\eta)\overline{v}}-1}
  \Big\rangle_h\;,
\end{eqnarray}
with $F_T$ the threshold force given in Eq.~(\ref{FT}) and $M$
given by
\begin{equation}
\label{slopeM}
M(\eta,K)=(1+\eta)\bigg\la\frac{h^2}{(K+h)^2}\bigg\ra_h\;.
\end{equation}

\begin{figure}
\centering
\includegraphics[width=12.cm]{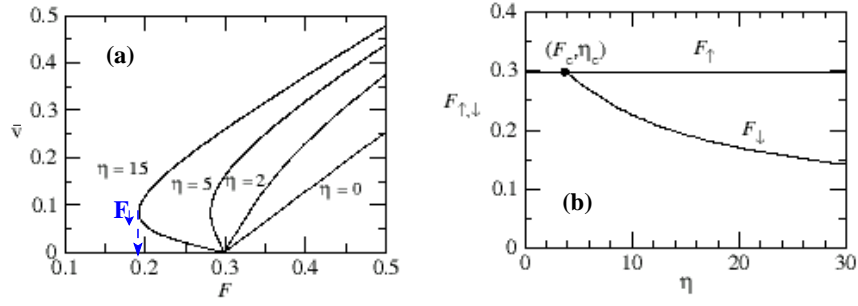}
\caption{\label{fig:VE_versus_eta} (a) Velocity versus driving
force for the VE model with $K=1$ and a broad distribution of
pinning strength, $\rho(h)=e^{-h}$. The velocity is continuous and
single-valued for $\eta<\eta_c$ and becomes multivalued for
$\eta>\eta_c$. The dashed line on he curve for $\eta=15$ indicates
the value  $F_\downarrow$ where the system repins when the drive
is ramped down from the sliding state.  Frame $(b)$ shows the
depinning and repinning forces $F_\uparrow$ and $F_\downarrow$ as
functions of $\eta$. The tricritical point at $(F_c,\eta_c)$
separates continuous from hysteretic depinning. Pinned and sliding
states coexist in the region $F_\downarrow<F<F_\uparrow$.}
\end{figure}

As in the purely elastic case ($\eta=0$) only static solutions
exist for $F<F_T$.  For $F>F_T$  there is a {\it unique} sliding
solution, provided $M(\eta,K)<1$, with mean velocity near
threshold given by
\begin{equation}
\label{MFvel_straight} \overline{v}\sim\frac{F-F_T}{1-M(\eta,K)}
\sim(1+\eta_c)\frac{F-F_T}{\eta_c-\eta}\;,
\end{equation}
giving $\beta=1$, as in the purely elastic case. The critical line
$\eta_c(K)$ separating unique from multivalued sliding solutions
is determined by $M(\eta,K)=1$,
\begin{equation}
\label{etac}
\eta_c(K)=\bigg\la\frac{h^2}{(K+h)^2}\bigg\ra_h^{-1}-1\;.
\end{equation}
The velocity-force curves and a phase diagram are shown in
Fig.~\ref{fig:VE_versus_eta} for $\rho(h)=e^{-h}$. There is  a
{\it tricritical point} at $(\eta_c,F_c=F_T)$. In contrast to the
purely viscous model with $K=0$, for finite long-time elasticity
($K>0$) the behavior is \emph{independent of the shape of the
pinning force distribution, }$\rho(h)$. For $\eta<\eta_c$, a
continuous depinning transition at $F_T$ separates a pinned state
from a sliding state with {\em unique} velocity.  In finite
dimensions, this transition is likely to remain in the same
universality class as the depinning of an elastic medium
($\eta=0$). In our mean-field example, the linear response
diverges at $\eta_c$, $v(\eta=\eta_c)\sim 1/\ln(F-F_T)$. For
$\eta>\eta_c$ there is hysteresis with coexistence of stuck and
sliding states.

Numerical simulations of the VE model in two dimensions
($d_\parallel=d_\perp=1$) indicate a strong crossover (possibly a
tricritical point) at a critical value of $\eta_c$ from continuous
to hysteretic depinning \cite{Bety_unp}. Although it is always
difficult to establish conclusively on the basis of numerics that
hysteresis survives in the limit of infinite systems, the size of
the hysteresis loop evaluated numerically does appear to saturate
to a finite value at large system sizes, indicating that the MF
approximation may indeed capture the correct finite-dimensional
physics.

\section{Relationship to other models and to experiments}
\label{SEC:other_models} Other models of driven systems with
inertial-type couplings have been proposed in the literature. It
is useful to discuss in some detail their relationship to the
viscoelastic model considered here.

In the context of charge density waves, Littlewood
\cite{littlewood88} and Levy and collaborators
\cite{levy92,levy94} modified the Fukuyama-Lee-Rice model
\cite{Gruner1988} that describes the phase of the CDW electrons as
an overdamped elastic manifold driven through quenched disorder by
incorporating the coupling of the CDW electrons to normal
carriers. This was realized via a global coupling in the equation
of motion for the phase to the mean velocity of the CDW, not
unlike what obtained by a  mean-field approximation of our viscous
coupling. The model was argued to account for the switching and
non-switching behavior observed in experiments.

Schwarz  and Fisher \cite{SF01,SF02} recently considered a model
of crack propagation in heterogeneous solids that incorporates
stress overshoot, that is  the fact that a moving segment of the
crack can sometimes overshoot one or more potential static
configurations before settling in a new one,  inducing motion of
neighboring segments. These effects may arise from elastic waves
that can carry stress from one region to another of the driven
medium. Stress overshoots, just like topological defects in a
driven disordered lattice, have an effect similar to that of local
inertia and were modeled by Fisher and Schwarz by adding couplings
to gradients in the local crack velocity in the equation of motion
for a driven elastic crack. These authors considered an automaton
model where time is discrete. It is straightforward  to define an
automaton version of our VE model, where both the displacement
$u_i$ and time are discrete, as shown in Ref.~\cite{MCMPramana}.
It is then apparent that the automaton version of the
viscoeleastic model given in Ref.~\cite{MCMPramana} is identical
in its dynamics to the model of crack propagation with stress
overshoot studied by Schwarz and Fisher, provided the strength $M$
of the stress overshoot is identified with the combination
$\eta/(1+\eta)$. The two models differ in the type of pinning
considered as the random force  used in by Schwarz and Fisher is
not periodic. We find, however, that the two models have identical
mean-field behavior, with a mean-field tricritical point
separating continuous from hysteretic dynamical transitions.  The
connection between the viscoelastic and the stress-overshoot model
is important because it stresses that distinct physical mechanisms
(inertia, nonlocal stress propagation, unbound topological
defects) at play in different physical systems can be described
generically by a coarse-grained model that includes a coupling to
local velocities of the driven manifold. Finally, in a very recent
paper, Maimon and Schwarz suggested that out of equilibrium a new
type of generic hysteresis is possible even when the phase
transition remains continuous \cite{Maimon2004}. Driven models
with both elastic and dissipative velocity couplings may therefore
belong to a novel universality class that exhibits features of
both first and second order equilibrium phase transitions. They
clearly deserve further study.

\begin{figure}
\centering
\includegraphics[width=10.cm]{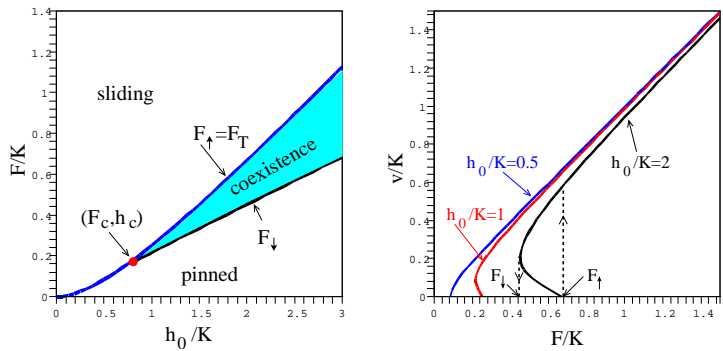}
\caption{\label{fig:VEPhaseD_F_vs_h} Mean-field solution of the VE
model with a piecewise parabolic pinning potential,
$\rho(h)=\delta(h-h_0)$ and $\eta=5$. Left frame: phase diagram.
Right frame: velocity versus drive for $h_0/K=0.5$ (blue),
$h_0/K=1$ (red) and $h_0/K=2$ (black). Also shown for  $h_0/K=2$
are the discontinuous hysteretic jumps of the velocity obtained
when $F$ is ramped up and down adiabatically.}
\end{figure}

We now turn briefly to simulations and experiments. For comparison
with experiments it is useful to point out that the tricritical
point of the viscoelastic model can also be obtained by tuning the
applied force and  the disorder strength, rather than the applied
force and the viscosity. Since the phase diagram does not depend
on the form of the disorder distribution, $\rho(h)$, we choose for
convenience a sharp distribution, $\rho(h)=\delta(h-h_0)$. The
phase diagram in the $(F,h_0)$ plane is shown in
Fig.~\ref{fig:VEPhaseD_F_vs_h}. For weak disorder the depinning is
continuous, while for strong disorder it becomes hysteretic, with
a region of coexistence of pinned and moving degrees of freedom.
The {\it tricritical point} is at $(h_c,F_c=F_T)$, with
$h_c=K/(\sqrt{1+\eta}-1)$.

Simulations of two-dimensional driven vortex lattices clearly show
a crossover as a function of disorder strength from an elastic
regime to a regime where the dynamics near depinning is spatially
inhomogeneous and plastic, with coexistence of pinned and moving
degrees of freedom
\cite{ShiBerlinsky1991,ReichardtOlsonSimul,GBD,Faleski1996}. In
fact a bimodal distribution of local velocity was identified in
Ref.~\cite{Faleski1996} as the signature of plastic depinning.
This local plasticity does not, however, lead to hysteresis in the
macroscopic dc response in two dimension: the mean velocity
remains continuous and single-valued, although it acquires a
characteristic concave-up ward form near depinning that cannot be
described by the exponent $\beta<1$  predicted by elastic models
in all dimensions. Hysteresis is, however, observed in simulations
of three-dimensional layered vortex arrays where the couplings
across layers are weaker than the in-layer ones
\cite{Olson_hyst01}. In this case the phase diagram is
qualitatively similar to that obtained for the viscoelastic model.

Recent experiments in $NbSe_2$ have argued that memory effects
originally attributed in this system to "plasticity' of the driven
vortex lattice \cite{Bhattacharya93}are actually due edge
contamination effects \cite{paltiel,Paltiel02,marchevsky}. In the
experiments a metastable disordered vortex phase is injected in a
stable ordered bulk vortex lattice. Memory effects may then arise
in the macroscopic dynamics during the annealing of the injected
disordered phase. Edge contamination does not, however, explain
the plasticity seen in simulations, where periodic boundary
conditions are used \cite{Faleski1996}. A possible scenario may be
that while in the experiments the vortex lattice in the bulk is
always in the ordered phase, in the simulations the vortex lattice
in the bulk of the sample may be strongly disordered even in the
absence of drive. Such a disordered vortex lattice would then
naturally respond plastically to an external drive. Finally, it is
worth mentioning one experimental situation where hysteresis of
the type obtained in our model is indeed observed in the
macroscopic response. This occurs in the context of charge density
waves, driven by both a dc and an ac field. In this case the dc
response exhibits mode-locking steps. The "depinning" from such
mode-locked steps was found to be hysteretic \cite{Higgins1993}.

\vspace{0.2in} Several colleagues and students have contributed to
various aspects of this work: Alan Middleton, Bety
Rodriguez-Milla, Karl Saunders, and Jen Schwarz. I am also
grateful to Jan Meinke for help with some of the figures. The work
was supported by the National Science Foundation via grants
DMR-0305407 and DMR-0219292.


\begin{thebibliography}{99.}
\bibitem{nonlocality_note}
The strain-stress relation is nonlocal in both space and time.
Only the nonlocality in time is retained here.
\end{thebibliography}
\end{document}